# A non-invasive device to measure mechanical interaction between tongue, palate and teeth during speech production

Christophe Jeannin[1,5], Pascal Perrier[1], Yohan Payan[2], André Dittmar[3], Brigitte Grosgogeat[4,5]

[1] Institut de la Communication Parlée, INPG, Grenoble, France
[2] TIMC, Grenoble, France
[3] Laboratoire des microsystèmes et microcapteurs biomédicaux, INSA, Lyon, France
[4] Laboratoire d'Etude des Interfaces et des biofilms en Odontologie, Lyon, France
[5] Hospices Civils de Lyon, Service d'odontologie, Lyon, France
Christophe.Jeannin@icp.inpg.fr, perrier@icp.inpg.fr

**Abstract**

This paper describes an original experimental procedure to measure the mechanical interaction between the tongue and teeth and palate during speech production. It consists in using edentulous people as subjects and to insert pressure sensors in the structure of their complete dental prosthesis. Hence, there is no perturbation of the vocal tract cavity due to the sensors themselves. Several duplicates are used with transducers situated at different locations of the complete denture according to palatography's results, in order to carefully analyze the production of specific sounds such as stop consonants.. It is also possible to measure the contact pressure at different locations on the palate for the same sound.

## I. INTRODUCTION

Speech motor control has been often compared with the control of other skilled human movements such as pointing or grasping [1]. This approach was very helpful and permitted the elaboration of important hypotheses that were the basis of major speech production theories. However, a peculiarity of speech movement has been often overseen, namely the fact that speech articulators, and especially the tongue, are not moving in a free space. Indeed, the vocal tract is a very narrow space, and tongue is most of the time in mechanical interaction with external structures, such as the palate or/and the teeth. Hamlet & Stone [2] and Fuchs et al. [3] have found a number of evidences supporting the hypothesis that these external structures would be integrated in speech motor control strategies, and would, consequently, significantly contribute to the control of speech movement accuracy.

Now, two questions can be raised:

(1) What is the quantitative nature of the interaction between tongue and external structures? In other words, are these structures only geometrical limits of the space in which tongue is allowed to move, or are they mechanical objects that are actually used to position and shape the tongue?

(2) What are the changes in speech motor control strategies induced by dramatic modifications of these external structures, as it is the case for instance for edentulous people?

Quantitative measurements of the intensity of the force exerted by the tongue on the teeth provide an interesting basis to address these issues. In addition, this technique provides interesting information about the order of magnitude of the intensity of muscular forces involved in the generation of tongue movements during speech production.

In this aim, a number of experimental set-ups have been developed in the past to measure tongue pressure against the palate in various experimental conditions [4]. The limits of these techniques, beside the inherent complexity of their calibration, lie in the fact that they actually induce slight perturbations of the speech production, because they modify the geometry of the vocal tract. Honda et al. [5] have shown that speakers can compensate quite easily and quite quickly for brutal changes in the thickness of an inflated palate, in that sense that they could adapt tongue positions in reference to the variable palatal shapes. However, it is not clear whether the intensity of the palate/tongue interaction was or not affected by these brutal changes.

In this paper, we will present a new experimental procedure that aims

(1) at measuring the interaction between tongue, teeth and palate without perturbing the production of speech, and

(2) at studying how speech motor control strategies evolve for edentulous people, from the moment where an artificial denture is put back in the mouth. Finally, preliminary results of a pilot study will be presented.

## II. EXPERIMENTAL DEVICE AND METHODS

The basic principle and the originality of the method presented in this paper is to use edentulous people as subjects, and to insert pressure sensors in their complete dental prosthesis, in such a way that the geometry of the vocal tract remains exactly the same as when their normal dental prosthesis is in place.

### A. Experimental device

General description

The complete denture, in which the pressure sensor is included, is placed inside the mouth. A sheath goes from the premolar area to the connector placed outside the mouth, via the labial commissure. Then, a wire goes from the connector to the amplifier, from the connector to a data sampling board and then to the computer.

A microphone is also connected to an amplifier to record the acoustic speech signal simultaneously with the pressure exerted by the tongue against the palate and/or the teeth. This amplifier is in turn linked to the data sampling board.

Description of the complete dental prosthesis

The dental prosthesis is made of resin and consists of complete artificial denture and of an artificial palate. Artificial teeth are similar in shape and size to natural teeth. The artificial palate must be at least 3mm thick to avoid breakage. Hence, both the pressure sensor and the wires connecting it with the connector outside of the mouth can be easily inserted in the prosthesis, without creating any additional change of the oral cavity (fig. 1).

For each edentulous patient, the dental prosthesis that is designed for obvious medical purposes is accurately duplicated thanks to a specific prosthesis design technique. Several duplicates are thus realized, in order to have different possible positions for the pressure sensor. Thus, the sensor can be inserted in the prostheses before the experiment with the patient, and no time is wasted during the experiment itself, when the tongue palate/teeth interaction is measured at different locations.

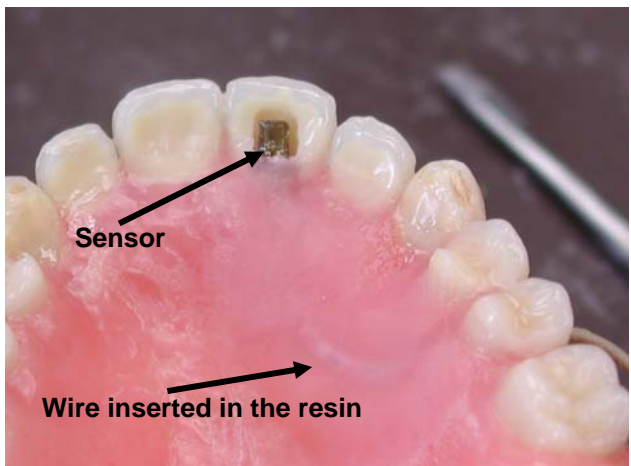

**Figure 1: Complete dental prosthesis with a sensor inserted in a front incisor. The wire connecting the sensor to the external connector can be seen on the right hand side of the picture, at the level of the first premolar. It goes then to the sensor inside the structure of the palate.**

Pressure sensor description

The sensor is made of a strain gauge sensor which is composed of thirteen layers (fig. 2). Each layer plays an important role in measuring capabilities of the transducer.

The intraoral area is a very difficult environment mainly because of three factors:
- Permanent moisture due to saliva
- Variable temperature
- Mechanical constrains

Therefore, the sensor must be electrically insulated and water proof. Moreover, it must be sturdy in order to go through several experiments.

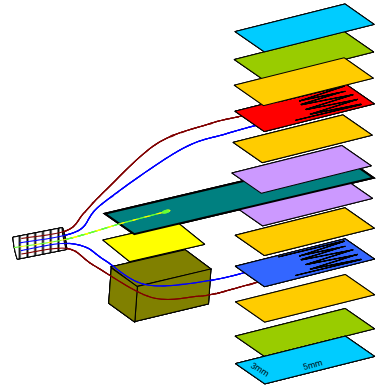

**Figure 2: The 13 layers of the pressure sensor**

The middle layer is made of a steel cantilever beam which is 10/100 mm thick. It supports on each side an active strain gauge. The gauges (Vishay, ref EA06 062 AQ 350) are placed in a half Wheatstone bridge configuration.. This strain gauge has been chosen because of its stability in temperature.

Gauges are bonded on a metallic support with M-Bond 200 adhesive (Vishay measurement group). Wires of 0,1mm diameter are soldered with tin on the gauge. The area of solder is 1mm$^2$ small and there are 4 points, 2 by side.

The two next layers are made of protective coating. M-Coat A (Vishay measurement group).

In any case, during the construction, the thickness of all the liquid components such as protective coating or bonding must be very small, in order to preserve the mechanical properties of the sensor. Consequently just one application of each component by layer can be done.

Figure 3 shows a detailed picture of such a sensor.

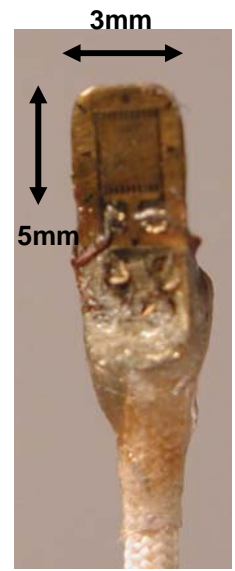

**Figure 3: The pressure sensor connected to its wire.**

Associated instrumentation tools:

Before sampling the signal goes through an amplifier (2100 series by Vishay Measurement) featuring a digital control display and different settings possibilities with 2100 maximal gain range and a bandwidth of 5kHz (at 0,5dB, and 15 kHz at -3 dB). It can hold up to two transducers.

To record the acoustic signal, a microphone is placed near the patient. It is linked with an amplifier and then with the data sampling board (DT 9800 series by Data translation) that is connected directly to the host computer via an USB port. This board can accept up to 16 analog inputs that can be simultaneously sampled at different rates (from 50 Hz to 20 000 Hz). For these experiments, 2 or 3 inputs are used: one for the acoustic signal and one or two for the tongue pressure.

## B. Methods

Pressure sensor calibration

Since the transducers are handmade, some differences can exist between them. Therefore, they have to be calibrated individually to convert electric signals into mechanical units such as strength or pressure.

The soft body characteristics of the tongue suggests that pressure should be more appropriate. Indeed, due to the tongue deformation in contact with a solid structure, the contact area is always large and can't be reduced to a specific point. However, the transformation of the sensor displacement into mechanical pressure is not obvious, because of the visco-elastic properties of the tongue. Indeed, in case of contact with external structures, the shape of the tongue varies over time and this variation is strongly dependent of the visco-elastic properties of the tongue. Hence, establishing a good approximation of the relation between the strain exerted on the sensor and the contact pressure is not a simple task. In this aim, we designed and tested 2 different devices to calibrate the sensors:

- The first device uses weights to convert electrical signals into strength.

This is the fastest and easiest way to calibrate the transducer. Small lead beads hanging out of the middle of the edge of the steel cantilever beam were used. Since the weight of the lead is known, the electric signal can be converted into mechanical strength. This allows an evaluation of the intensity of the mechanical interaction, as well as a comparison of them under different experimental conditions. However, it is not a good approach to quantitatively asses their absolute values of the pressure at contact location.

- The second device called "dried water column" converts electric signals into pressure.

The weight of a water column is applied on the whole surface of the sensor. A latex membrane which is not tensed (to avoid signal due of it) is attached to the end of the column and is in contact with the sensor. The electric signal is compared to the level of water. This is a nice way to account for the soft body characteristics of the tongue, and to give an idea of the pressure at contact location. However, it does not model the true viscoelastic characteristics of the tongue. Consequently, the conversion from sensor displacement to contact pressure does not strictly apply to the contact between tongue and teeth and palate.

In both cases, the calibration of the sensors has to be done very carefully and to be explained in order to know what kind of information can be extracted from the data.

Palatography

In order to know with enough accuracy where the sensors should be inserted to measure tongue-palate/teeth interaction during speech production, the exact locations of the main contact regions have to be determined. [6]. This is why, during a preliminary session, palatographic recordings are carried out, with the prosthesis in the mouth

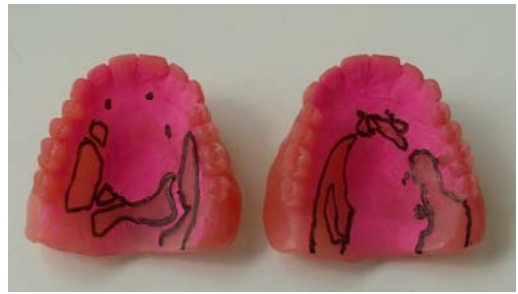

**Figure 4: Tongue contacts regions on the prosthesis obtained with palatography for /t/ (left) and /d/ (right), and for a female subject in a pilot experiment.**

With complete dental prostheses, electropalatagraphy (EPG) can not be used because of obvious incompatibility reasons between the prosthesis and the EPG device [7]. Therefore, pink powder (occlusion spray red "okklufine premium™") is applied on the teeth and on the artificial palate. When tongue is in contact with one of these structures, the powder is removed from the area of contact. Thus, when the subject is asked to pronounce a specific phoneme in isolation, the edges of the contact areas can be determined for this phoneme. These edges are highlighted with a black pen before removing the powder. This technique is not as accurate as

EPG, but accurate enough to determined where to set the sensor when tongue and palate/teeth interaction is investigated for this phoneme.

Figure 4 shows an example of the results thus obtained during the production of the alveolar stops /t/ and /d/, for a female subject in a pilot experiment.

Data acquisition

As exemplified above, for each subject, each prosthesis is dedicated to the measurement of tongue – palate/teeth interactions in a vocal tract region that is strongly related to the production of a specific phoneme. Sounds are repeated several times by the subjects both in isolation and within short carrier sentences such as, for the alveolar stop /t/, "toto a têté sa tétine".

For each sound, according to the palatography results, the transducers are placed in the specified area.

## III. RESULTS

Figure 5 shows an example of results obtained for /d/ in a pilot study carried out with an 80 years old female subject. The sensor was inserted in the most front contact area measured with palatography (see fig. 4, right panel). It can be seen that the acoustic release of the stops is well-synchronized with the abrupt decrease of tongue pressure in the palate. The vertical axis represents the intensity of force exerted by the tongue on the steel cantilever.

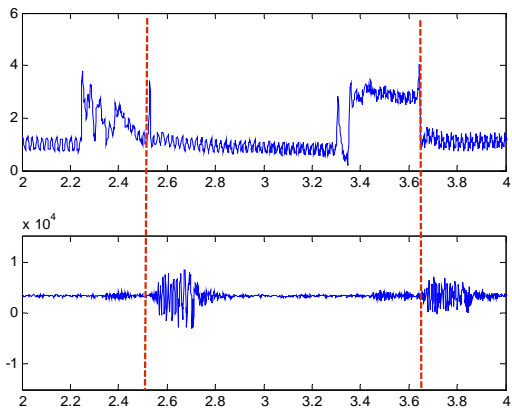

**Figure 5: Acoustic speech signal (low panel) and tongue force in the alveolar region (upper panel) during two repetitions of [de]**

Obviously, the pressure patterns depict a noticeable variability. The origin of this variability has to be clarified, to know whether it is related to the experimental device or whether it reveals the intrinsic intra-speaker variability of speech production. The maximal order of magnitude of the force is around 0.003N which is in agreement with other kind of data published in the literature.

## IV. CONCLUSION

An original device for the measurement of the mechanical interaction between tongue and teeth and/or palate was presented. It adapted to a specific kind of subjects, namely edentulous patients. Using the complete dental prosthesis to insert force sensors, the device permits the measurement of contact pressure without introducing any additional perturbation than the prosthesis it self.

This experimental setup will permit to study speech production either by patients who have been wearing their prosthesis for years and have completed the adaptation process to it, or by patients that just received the prosthesis, in order to study how they adapt to its new denture..